# Correlation enhanced resistance hysteresis near half filling in $MoS_2/WSe_2$ heterobilayer

**Authors:** Yong Chen(陈勇)[1, 2#], Weikang Zhang(张炜康)[1#], Meizhen Huang(黄美珍)[1#*], Shengling Xiang(项圣凌)[1, 2], Zishu Zhou(周子澍)[1], Yaqi Ma(马亚琪)[1], Chenxuan Lou(楼晨煊)[1], Haoxi Ji(姬浩曦)[1], Aoqian Zhang(张傲乾)[1], Yifei Jin(金逸飞)[1], Liheng An(安礼恒)[1], Zefei Wu(吴泽飞)[1], Chun Cheng(程春)[2*], and Ning Wang(王宁)[1*]

**Affiliations:**

[1]Department of Physics and State Key Laboratory of Optical Quantum Materials, the Hong Kong University of Science and Technology, Kowloon, Hong Kong, China.

[2]Department of Materials Science and Engineering, Southern University of Science and Technology, Shenzhen 518055, China.

[#]These authors contributed equally to this work.

*Correspondence to: mhuangai@connect.ust.hk; chengc@sustech.edu.cn; phwang@ust.hk.

## Abstract

Ferroelectricity, typically arising from ionic displacements in noncentrosymmetric lattices, enabling applications in memory devices and sensors. Recent advances in two-dimensional materials and van der Waals heterostructures have revealed novel ferroelectric phenomena, including sliding ferroelectricity and correlation-driven ferroelectricity in moiré superlattices. In this work, we fabricate and study a $MoS_2/WSe_2$ moiré superlattice device exhibiting a high field-effect mobility of 17,650 $cm^2V^{-1}s^{-1}$. Electrical transport measurements reveal correlated insulating states accompanied by a prominent and reproducible resistance hysteresis near half filling. Temperature and displacement field dependence further confirms the correlation-enhanced nature of the hysteresis. Our analysis suggests that displacement field-induced metal-to-insulator transition at correlated insulating state coupled with interfacial dipoles enables the observed resistance hysteresis. These results establish correlation enhanced resistance hysteresis near half filling in a $MoS_2/WSe_2$ heterobilayer, offering opportunities for exploring emergent quantum phases and device functionalities.

## *1. Introduction*

Ferroelectricity, defined by the presence of spontaneous and reversible electric polarization, has long been vital in condensed matter physics and material science. In traditional systems like perovskite oxides (e.g., $BaTiO_3$, $PbTiO_3$)[1-3], ferroelectricity arises from collective ionic displacements in noncentrosymmetric lattices, producing a macroscopic polarization that can be switched by an external electric field. This classic understanding has driven the development of applications such as non-volatile memory, sensors, and actuators[4-10]. Recent advances in two-dimensional (2D) materials and van der Waals (vdW) heterostructures have dramatically expanded the landscape of ferroelectricity[11-16], giving rise to novel mechanisms such as sliding ferroelectricity, in which polarization arises from interlayer sliding rather than internal atomic or lattice shifts. This effect has been observed in materials like few-layer $WTe_2$[17-19], twisted hexagonal Boron Nitride (hBN)[20-22], and various twisted transition metal dichalcogenide (TMDC) heterostructures[23, 24]. These systems enable electrical control of polarization through mechanisms inaccessible in bulk materials.

In parallel, correlation-driven ferroelectricity has emerged in moiré superlattices formed by stacking 2D crystals at small twist angles[25-31]. Here, flat electronic bands enhance interactions, leading to rich correlated physics including Mott insulating states[25, 32-34] and unconventional superconductivity[35-37]. Ferroelectric order can also develop from these strong correlations, as seen in twisted $WSe_2$[29] and multilayer $MoS_2$[27], where ferroelectric and correlated insulating states coexist. Despite exciting discoveries, the microscopic origins of correlation-driven ferroelectricity remain uncertain, with ongoing research into mechanisms such as electronic nematicity, charge order, and symmetry breaking. This evolving field offers both challenges and opportunities for uncovering new quantum phases[38] and functional devices[6, 7, 9].

Here, we fabricate and investigate $MoS_2/WSe_2$ moiré superlattice device with a high field-effect mobility of 17,650 $cm^2V^{-1}s^{-1}$. Electrical transport measurements reveal prominent insulating states at half and full moiré band fillings, indicating the formation of flat bands and correlated electronic phases. Notably, near half filling we observe a robust ferroelectric signature in transport, manifested as a pronounced resistance hysteresis under gate sweeps. This resistance hysteresis persists over a wide range of scan rates, is progressively suppressed with increasing displacement field and temperature, and disappears above 50 K. Importantly, its behavior closely tracks the correlated insulator state, indicating a correlation-dependent origin distinct from conventional or sliding ferroelectrics. We find that a displacement-field-tuned metal-insulator transition at the correlated insulating state may contribute to the observed pronounced resistance hysteresis.

## 2. *Resistance hysteresis in high-mobility moiré devices*

Figs. 1(a) and 1(b) show the schematic view and the optical image of the high mobility dual-gated $MoS_2/WSe_2$ moiré device. The device is constructed by encapsulating a $MoS_2/WSe_2$ heterobilayer between two flat hBN flakes using dry-transfer technique The two monolayer TMDCs, exhibiting a lattice mismatch of approximately 4%, are stacked in parallel to form a triangular moiré superlattice with a lattice constant $L_{\mathrm{M}} \sim a_{\mathrm{WSe_2}}/\delta \sim 8\ \mathrm{nm}$, where $a_{\mathrm{WSe_2}} =$ $0.328$ nm and $a_{\mathrm{MoS_2}} = 0.315$ nm are the lattice constants of $WSe_2$ and $MoS_2$, respectively, and $\delta = |a_{\mathrm{MoS_2}} - a_{\mathrm{WSe_2}}|/a_{\mathrm{MoS_2}} \approx 4\%$ is the relative lattice constant mismatch[39]. This 8 nm moiré length gives rise to a moiré full filling density of $2n_0 = 4/(\sqrt{3}L_{\mathrm{M}}^2) \sim 3.8 \times 10^{12}\ \mathrm{cm}^{-2}$.

To probe the transport properties, we fabricate the electrical contacts using our previously invented local bonding distortion strategy (detailed fabrication process can be found in Methods section and Supplementary Fig. S1)[40]. A dual gate geometry is applied to allow independently control of the external out-of-plane displacement field $D = (C_{\mathrm{bg}}V_{\mathrm{bg}} - C_{\mathrm{tg}}V_{\mathrm{tg}})/2\varepsilon_0$ and the carrier density $n = (C_{\mathrm{bg}}V_{\mathrm{bg}} + C_{\mathrm{tg}}V_{\mathrm{tg}})/e + n_{\mathrm{res}}$, where $\varepsilon_0$ is the vacuum permittivity and $e$ is the elementary charge, $C_{\mathrm{bg}}$ and $C_{\mathrm{tg}}$ are the geometric gate capacitances per area between the respective gate and the $WSe_2$/$MoS_2$ flakes. Here $n_{\mathrm{res}}$ denotes the residual background carrier density, accounting for charge-offset doping from impurities/trapped charges and shifting the density zero point in the capacitor model. We patterned the device into Hall-bar geometry and measured the four-probe longitudinal resistance $R$ down to 2.0 K. As illustrated in Fig. 1(c), it reveals a high FET mobility of 17,650 $\mathrm{cm^2V^{-1}s^{-1}}$. Detailed extraction of these parameters can be found in Methods and Supplementary Note 1.

We systematically investigate the electrical transport properties of the device through dual-gate modulation. As shown in Fig. 1(d), the resistance curves under top-gate sweeping at various bottom-gate voltages each exhibit two pronounced peaks, according to the determined carrier density via Hall measurements (details can be found in Supplementary Note 1), we find these peaks corresponding to the one electron per moiré unit cell (half-filling state) and two electron per moiré unit cell (full-filling state). The emergence of these insulating states offers direct evidence for flat-band formation and correlated electronic phases in the $MoS_2$/$WSe_2$ heterostructure. Furthermore, clear hysteresis loops are observed between forward and backward sweeps (marked by black arrows). Here, we define the forward sweep as increasing $V_{\mathrm{tg}}$ from negative to positive values, which (at fixed $V_{\mathrm{bg}}$) corresponds to a decrease in the applied displacement field $D$; the backward sweep reverses $V_{\mathrm{tg}}$ from positive back to negative (corresponding to increasing $D$). This convention is used throughout the manuscript for all hysteresis loops. The robustness of this hysteretic response was confirmed through repeated measurements across a range of scan rates and excitation amplitudes (see Supplementary Note 2 for details). The persistence of the hysteresis under varying scan rates rules out charge trapping, which would be rate-dependent, while its insensitivity to excitation amplitudes together with the linear two-probe *V-I* characteristic excludes contact barrier effects. In addition, a series of further experiments (see Supplementary Note 2) were performed to eliminate other possible artifacts such as gate leakage. Taken together, these results demonstrate that the hysteresis arises from intrinsic collective electronic behavior in the moiré superlattice.

### *3. D-field-dependent resistance hysteresis*

The observed ferroelectric-like transport signature fundamentally differs from conventional and sliding ferroelectrics. The primary distinction involves its exclusive emergence near half filling, where resistance hysteresis develops while remaining absent at band edges. This filling-dependent behavior contrasts sharply with conventional ferroelectrics' uniform polarization response. The exact coincidence between the resistance hysteresis and half filling instead points to an intrinsic coupling between the hysteretic transport response and the correlated electronic state.

To understand the electronic characteristics, we test the density-dependent transport under opposite sweeping directions. Figures 2(a) and 2(b) show the four-terminal longitudinal resistance $R_{\text{for}}$ and $R_{\text{back}}$ as a function of displacement field and carrier density during forward ($+D \rightarrow -D$) and backward ($-D \rightarrow +D$) field scans, respectively. $R_{\text{for}}$ shows two well-defined peaks appear at $n = n_0$ and $n = 2n_0$ [Fig. 2(a)], confirming half and full filling of the moiré superlattice (moiré filling factors $v = 1$ and 2). In contrast, under backward sweep [Fig. 2(b)], the peak at full filling $v = 2$ remains robust, whereas the feature near half-filling $v = 1$ is strongly suppressed. This pronounced sweep-directional asymmetry further indicates a link between correlated insulating states and the resistance hysteresis.

We quantify the hysteresis strength via the relative resistance difference $\Delta R/R_{\text{for}} = (R_{\text{for}} - R_{\text{back}})/R_{\text{for}}$ and perform a mapping of $\Delta R/R_{\text{for}}$ in Fig. 2(c), it shows a sharp peak at $v = 1$ and decreases as density deviates from half-filling, vanishing completely at band edges (see Fig. S10). Fig. 2(d) shows the hysteresis and the calculated strength at half filling state. Increasing displacement field $D$ suppresses the relative resistance difference, indicating a reduced hysteresis strength at elevated fields. This reduction can originate from both conventional ferroelectricity and a diminished stability of the correlated state under elevated fields. First, in conventional ferroelectrics, for a fixed electric field sweep range, increasing the magnitude of the applied field tends to narrow the polarization difference ($\Delta P$) observed in the $P$-$E$ hysteresis loop. As a result, the change in resistance and so the hysteresis strength, which should be proportional to the ferroelectric polarization, also becomes smaller. Second, electric-field modulation of the electronic band structure directly disrupts charge correlations, resulting in a sharp decrease in the correlated gap[41, 42] and forward-scan resistance $R_{\text{for}}$, as clearly evidenced by the rapid drop in the black dotted curve in Fig. 2(d). This suppression of electronic correlations propagates to the resistance hysteresis through their coupling, consequently suppressing hysteresis strength. The detailed physical mechanisms underlying this dual suppression will be analyzed in subsequent sections.

### *4. Temperature-dependent hysteretic transport behavior*

Considering that the correlated insulator state is suppressed at elevated temperatures[25], the hysteretic transport is also expected to exhibit temperature sensitivity. To investigate this, we measured the hysteresis loops at various temperatures [Fig. 3(a)]. Both the half- and full-filling resistance peaks gradually diminish as temperature increases. At high temperatures, the thermal energy provided to the electrons disrupts the conditions described by the Hubbard model, causing on-site Coulomb repulsion to no longer dominate the electronic behavior. Consequently, the half-filling peak vanishes first upon heating. A key observation is that the hysteresis loop shrinks concurrently with the disappearance of the half-filling signature. The hysteretic response collapses completely below 50 K, which lies well within the temperature range where the electron-electron correlation effect vanishes and is notably lower than the structural transition temperatures of typical sliding ferroelectrics such as $WTe_2$ ($> 300$ K)[17].

We further performed a quantitative analysis of the temperature-dependent resistance. As shown in Fig. 3(b), the forward- and backward-scan resistance values at half-filling exhibit distinct behavior: the backward-scan resistance $R_{\mathrm{back}}$ (blue dotted curve) increases with temperature, indicating metallic behavior, while the forward-scan resistance $R_{\mathrm{for}}$ (red dotted curve) decreases below 8 K, consistent with a correlated insulator state. Around 8 K, the system undergoes an insulator-to-metal transition. A thermal activation fit yields an energy gap of $\Delta = 0.38 \pm 0.03$ meV, corresponding to a characteristic activation temperature of $T_{\mathrm{c}} \approx 4.4$ K (inset). The emergence of insulating behavior below ~ 8 K agrees well with this energy scale, as thermal excitation across the gap becomes suppressed when $T \lesssim 2T_{\mathrm{c}}$.

Simultaneously, the normalized resistance difference $\Delta R/R_{\mathrm{for}}$ can be fitted well by $\Delta R/R_{\mathrm{for}} \sim \exp(-T_0/T)$ with $T_0 = 14.1$ K [Fig. 3(c)]. This temperature-driven suppression aligns with the behavior observed in correlated electronic states and suggests that the hysteresis strength decays exponentially with increasing temperature. This behavior is consistent with a correlation-driven origin: as thermal fluctuations weaken the correlated insulating state, the hysteresis correspondingly diminishes, supporting our description of the hysteresis as correlation-enhanced.

### *5. Possible mechanisms for the correlation enhanced hysteretic transport*

So far, we have demonstrated the $\nu$-, $D$-, and $T$-dependent resistance hysteresis, the co-enhancement of $\Delta R/R_{\mathrm{for}}$ and the insulating-state strength in both measurements provides strong evidence for the "correlation-enhanced" hysteresis (a quantitative relation between hysteresis strength and the insulating-state strength as functions of $\nu$, $D$ and $T$ is provided in Supplementary Note 3 to prove the correlation-enhanced nature).

Here, we propose a possible mechanism behind the emergence of resistance hysteresis in the moiré $MoS_2/WSe_2$ system. Analogous to other moiré systems[21, 25, 43], parallel-stacked heterobilayers exhibit inversion symmetry breaking, generating localized out-of-plane polarization. As depicted in Figs. 4(a) and 4(b), adjacent R-stacked domains host antiparallel polarizations ($P\uparrow/P\downarrow$) with atomically distinct origins, where $P\uparrow$originates from W-S interfacial dipoles in $AB_{Se}$ regions, and $P\downarrow$ derives from Mo-Se dipoles in $AB_W$ regions[32]. Within twisted $MoS_2/WSe_2$ moiré superlattices, these polarized domains constitute a bistable system. The polarization switching is then governed by a sliding mechanism, where an external vertical electric field drives a lateral relative displacement between the constituent layers to toggle between different stacking states with opposite dipole moments[22]. This process yields hysteretic *P*-*D* loops characterized by bistable polarization states at fixed electric field *D* [Fig. 4(c)]. Within this mechanism, the ferroelectric signal decreases with increasing domain size, which is controlled by twist angle or lattice mismatch[44, 45]. This behavior originates from the twist-angle-dependent structural transition of the moiré superlattice: at small twist angles (large domain size), commensurate triangular domains with soft, switchable domain walls enable ferroelectricity; above a critical angle (small domain size), a rigid hexagonal incommensurate pattern appears, suppressing ferroelectric switching. The 4% lattice mismatch for our heterobilayer system confines characteristic domain sizes in the moiré superlattice to scales smaller than its period ($\sim 8$ nm). This geometric constraint suppresses ferroelectric behavior. Nevertheless, this weak polarization modifies the effective displacement field within the system such that $D_{\rm eff} = D - P/\epsilon_0$, where $D$ is the external electric field and $P$ is the internal polarization, as illustrated in Fig. 4(d).

We propose that the displacement field induced metal to insulator transition can enhance the hysteretic behavior. Figure 4(e) exhibits resistance hysteresis under dual-gate modulation with fixed $V_{\rm bg} = 73$ V. Schematic band structures at those special states labelled from i to iv are shown in Fig. 4(f). During forward scanning ($V_{\rm tg}: -2 \rightarrow +3.5$ V), the system evolves through distinct states. State I: Initial high-resistance insulation at $V_{\rm tg} = -2$ V and $n \sim 0$ persists under strong displacement field $D \sim 0.35$ V/nm with dominant $P\uparrow$ polarization. State II: Resistance collapses sharply to ~3 kΩ at $V_{\rm t} = -0.2$ V as electrons delocalize with weakening fields. A distinct peak emerges at $V_{\rm tg} = 0.11$ V, corresponding to a correlated charge ordering near half-filling where moderate field $D = 0.14$ V/nm triggers partial $P\uparrow \rightarrow P\downarrow$ switching with Mo-Se dipoles stabilized. State III: Further scanning establishes insulating state at full filling ($V_{\rm tg} = 2$ V, $D = -0.042$ V/nm) with complete $P\downarrow$ dominance.

Reverse scanning reveals distinctive dynamics: The bistable polarization in hysteretic *P*-*D* loops causes the system to experience a different modified displacement field $D_{\rm eff}$ during backward scans compared to the forward sweep. Consequently, when returning to the same gate voltage (state IV), the effective field $D_{\rm eff}$ exceeds its value during forward scanning. When this effective field exceeds a critical threshold, dielectric screening suppresses electronic correlations, progressively narrowing the Hubbard gap until spectral overlap occurs. As the Hubbard gap decreases with increasing electric field[41, 42], this directly drives the observed resistivity reduction. At other filling factors, smaller resistance variations with electric field yield weaker ferroelectric-like transport signals. Therefore, the $D$ field-induced metal-to-insulator transitions enhance hysteretic behavior, qualitatively explaining the observed characteristics. We emphasize that the above mechanism is a qualitative scenario consistent with our experimental observations and remains a possible picture. Further verification will require further experimental and theoretical efforts.

## *6. Conclusion*

This work demonstrates correlation-enhanced resistance hysteresis in a high-mobility $MoS_2$/ $WSe_2$ moiré heterostructure. Crucially, resistance hysteresis emerges exclusively near half-filling ($\nu = 1$) and vanishes at the band edges, indicating a strong linkage between the hysteresis strength and the presence of correlation-driven electronic states. Mechanistically, interfacial dipoles ($P\uparrow$ in $AB_{Se}$, $P\downarrow$ in $AB_W$) create bistable domains, where lattice mismatch confines polarization scales ($< 8$ nm), suppressing sliding ferroelectricity but enabling effective field modulation via $D_{\mathrm{eff}} = D - P/\epsilon_0$. This framework explains how metal-insulator transitions at critical $D_{\mathrm{eff}}$ amplify hysteresis through Hubbard gap manipulation, as spectral overlap reduces resistivity during reverse scans. Overall, our results establish a correlation-enabled route to bistable, history-dependent transport in a moiré semiconductor platform, and highlight dual electrostatic control as a promising knob for designing low-power electronic memory-like functionalities based on correlated states.

## Methods

### *1. Sample preparation*

$MoS_2$/$WSe_2$ heterostructure device was fabricated through a dry pickup-transfer process initiated by mechanical exfoliation of $MoS_2$ and $WSe_2$ flakes onto $SiO_2$/Si substrates (285 nm oxide), with flake identification via optical contrast microscopy. Target flakes were subsequently encapsulated using a PDMS/PC polymer stamp assembly to sequentially transfer hBN bottom gates, selected $WSe_2$ and $MoS_2$ layers, and additional hBN top dielectric layers, followed by thermal annealing at 300°C for 3 hours in a $H_2$/Ar atmosphere to eliminate interfacial contaminants. Ohmic contacts are fabricated by our previous develop strategy[40] (see Supplementary Fig. S1 for details). Device isolation into Hall bar configurations was ultimately achieved through high-rate reactive ion etching using a $CHF_3$/$O_2$ mixture (40:4 sccm).

### *2. Electrical measurement*

Electrical measurements were performed in a cryogenic system with a base temperature of 2.0

K. The back gate voltage was provided by an Aim-TTi International PLH120 DC power supply. A Stanford Research DS360 low-distortion function generator was used to apply a 4.79 Hz source-drain bias voltage. The current was measured by a Signal Recovery 7280 wide-bandwidth digital lock-in amplifier, and the longitudinal and Hall voltages were measured by Stanford Research SR830 lock-in amplifiers with SR550 voltage preamplifiers. All top gate sweeps were carried out using a staircase step-and-settle protocol: the gate voltage is initially set to the negative maximum to fully polarize the ferroelectric layer in one direction before the sweep begins, then, sweep from a negative maximum to a positive maximum and then back to the negative maximum is performed; during the sweep, the top-gate voltage was incremented by $\Delta V_{tg}$ = 0.01 V per step, followed by a 1 s settling time unless other stated. Lock-in then readout data used a time constant of 1 s, a low-pass filter slope 18 dB/oct, with LINE and 2×LINE notch filters enabled (reserve mode: Normal). No numerical smoothing was applied in post-processing.

### *3. Extraction of the carrier density, displacement field, mobility and other parameters*

We extract the gate capacitances directly from Hall carrier density slopes. Specifically, we measured $R_{xy}(B)$. To suppress the admixture from the longitudinal component $R_{xx}$ due to contact misalignment, we antisymmetrized the Hall resistance as $R_{xy}(B) = [R_{xy}(+B) - R_{xy}(-B)]/2$ at $B$ = 1 T. The Hall density is then obtained from the low-field slope. And from the linear dependence of $n_{Hall}$ on gate voltage, we extract the gate capacitances directly.

We define $n_0$ as the carrier density corresponding to half filling of the moiré band (i.e., the density at which $v$ = 1). With this convention, full filling corresponds to $v$ = 2. To extract $n_0$, we use the top-gate calibration and determine the average top-gate voltage spacing between the experimentally identified half-filling and full-filling features.

The field-effect mobility is calculated using $\mu_{FE} = \frac{dG}{dV_{bg}} \frac{L}{W} \frac{1}{C_{bg}}$. The conductance $G$ used in Fig. 1c is the four-probe channel conductance which minimizes contact-resistance effects. We extract the transconductance from a linear fitting window of the $G$-$V_{bg}$ trace and then calculate the mobility. We further provide a Hall-mobility cross-check.

Detailed extraction of these parameters can be found in Supplementary Note 1.

**Acknowledgements**

Grant support from the National Key R&D Program of China (2020YFA0309600) and the Research Grants Council (RGC) of Hong Kong (Project Nos. 16303123, C6053-23G-D, AoE/P701/20) are acknowledged. We also acknowledge Mr. Chun Kit Lai, Mr. Gordon C.T. Suen, and Dr. Yuan Cai for their valuable technical support during device fabrication at the MCPF and WMINST of HKUST.

**Contributions**

N. W., M.-Z. H., Y. C., and W.-K. Z. conceived and designed the study. Y. C. prepared the samples. W.-K. Z., M.-Z. H. and Y. C. performed transport measurements. S.-L. X. conduct TEM measurement, M.-Z. H., W.-K. Z. and Y. C. analyzed data and wrote the manuscript with substantial input from N. W.. Z.-S. Z., Y.-Q. M., C.-X. L., H.-X. J., A.-Q. Z., Y.-F. J., L.-H. A., Z.-F. W., and C. C. provided technical support in the device fabrication process. N. W. finalized the manuscript with contributions from all authors.

**Supplementary Material**

Supplementary Materials are available, with the following contents:

Fig. S1. The RIE etching process for top-layer contact.

Note. 1. Extraction of the carrier density, displacement field, mobility and other parameters.

Note. 2. Rule out of extrinsic electrostatic artifacts.

Note. 3. Correlation-enhanced hysteresis: quantitative relation to insulating-state strength.

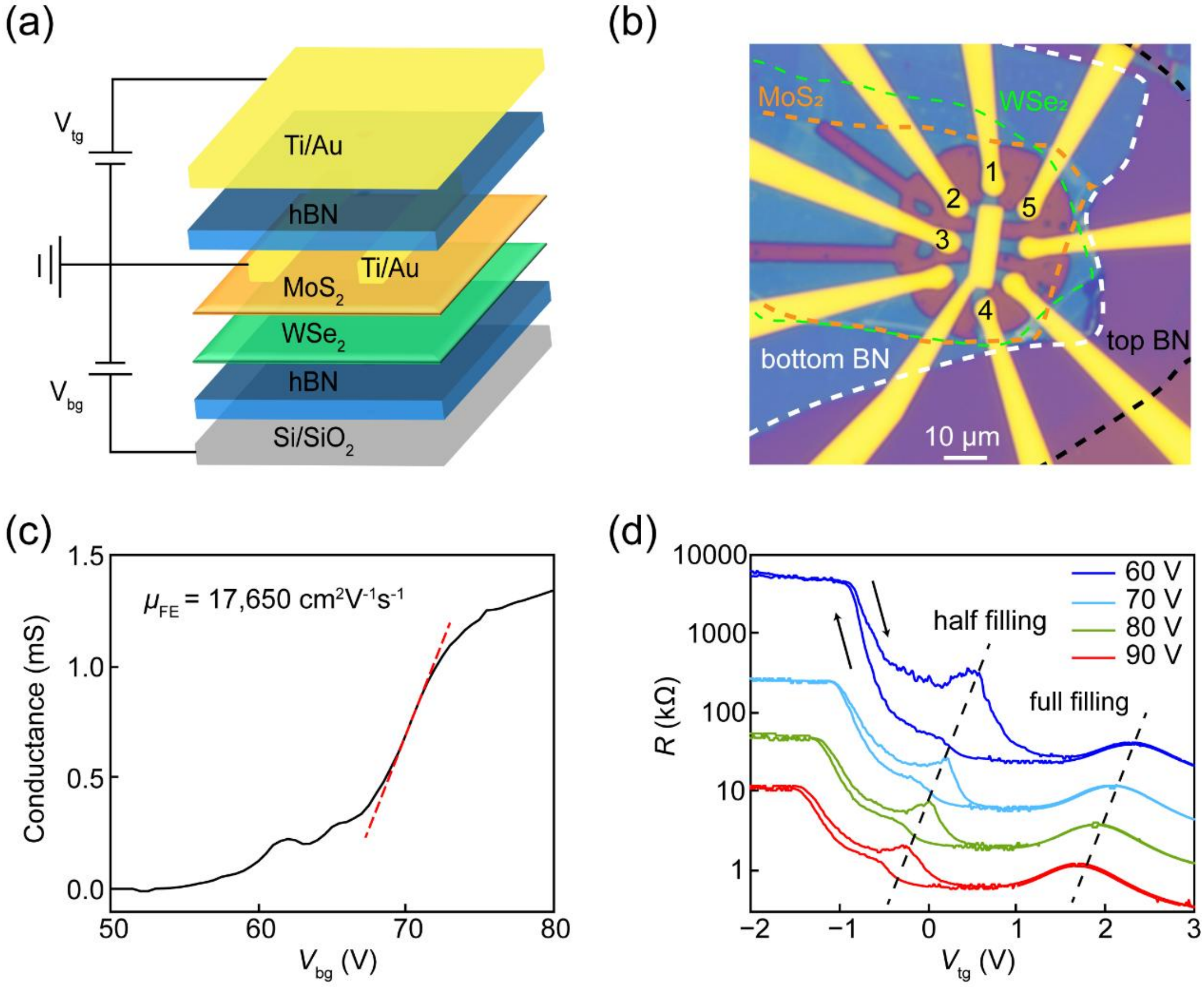

Fig. 1. (a) Schematic view of the device structure. (b) Optical image of the device. The dashed lines indicate each 2D layer. The current is driven between electrodes 1 and 4, while $R$ is obtained from the voltage drop measured between electrodes 2 and 3. (c) Gate dependent conductance at $T = 2.0$ K and top gate grounded. The red linear fit line indicates a high field-effect mobility of 17,650 $cm^2V^{-1}s^{-1}$. (d) Resistance hysteresis near half-filling at different bottom gates. The black arrows indicate the sweep direction (forward: $V_{tg}$ swept from negative to positive; backward: $V_{tg}$ swept from positive to negative). The black dash lines indicate the states of half and full filling.

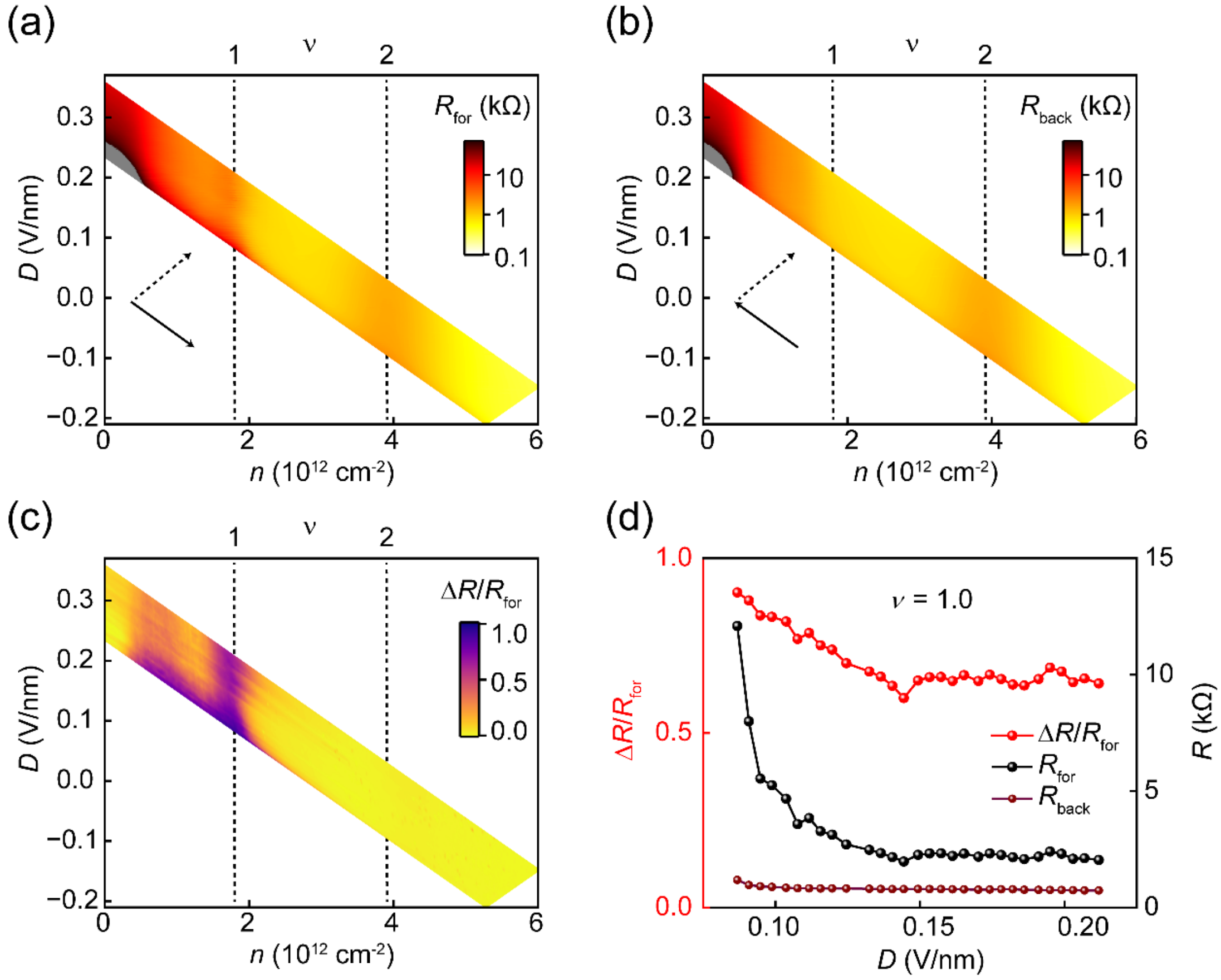

Fig. 2. (a), (b) Four-terminal resistance as functions of carrier densities and displacement fields for forward (a) and backward (b) scanning. We repeatedly scanned $V_{tg}$ (fast scan, solid arrow) in the both forward and backward direction while gradually changing $V_{bg}$ (slow scan, dashed arrow). The insulating states are marked by vertical dashed lines. Resistance peak at $v = 1$ is less pronounced under backward scanning conditions. (c) Hysteresis mapping quantified by the relative resistance change $\Delta R/R_{\mathrm{for}} = (R_{\mathrm{for}} - R_{\mathrm{back}})/R_{\mathrm{for}}$. The maximum hysteresis strength occurs at half-filling and decreases with increasing $D$ field. (d) Resistance and the calculated hysteresis strength at half filling state versus $D$ field. The black and dark red dotted curves denote the resistance measured during forward and backward sweeps, respectively, while the red dotted line represents the relative resistance change $\Delta R/R_{\mathrm{for}}$.

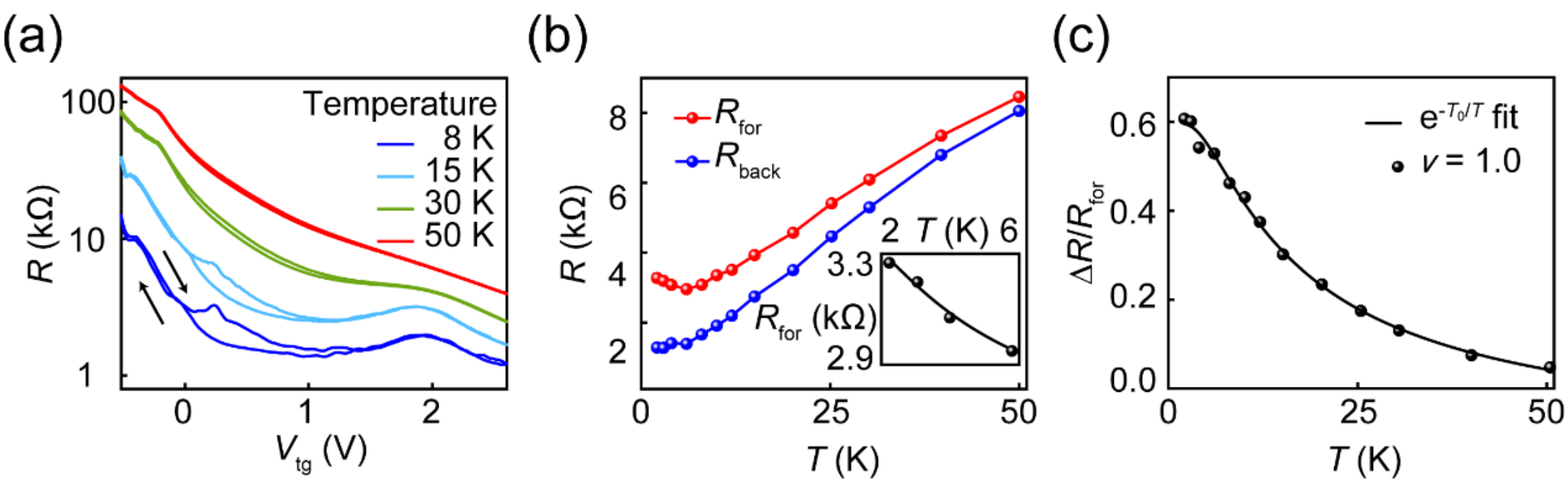

Fig. 3. (a) Temperature dependent hysteresis at $V_{\mathrm{bg}} = 75\,V$. The black arrows indicate the sweep direction. (b) Temperature dependent resistance at $\nu = 1$ measured under forward and backward sweeps. The resistance exhibits insulating behavior below 8 K for the forward sweep but a metallic behavior for backward sweep. Inset: Thermal activation fit of $R_{\mathrm{for}}$ versus temperature in the correlated insulator state, yielding an energy gap of $\Delta \approx 0.38 \pm 0.03$ meV and activation temperature of $T_{\mathrm{c}} = \Delta / k_{\mathrm{B}} = 4.4$ K. (c) Temperature dependence of hysteresis strength at $\nu = 1$. The solid line indicates exponential decay fit $\Delta R / R_{\mathrm{for}} \sim \exp(-T_0/T)$ with $T_0 \sim 14.1$ K.

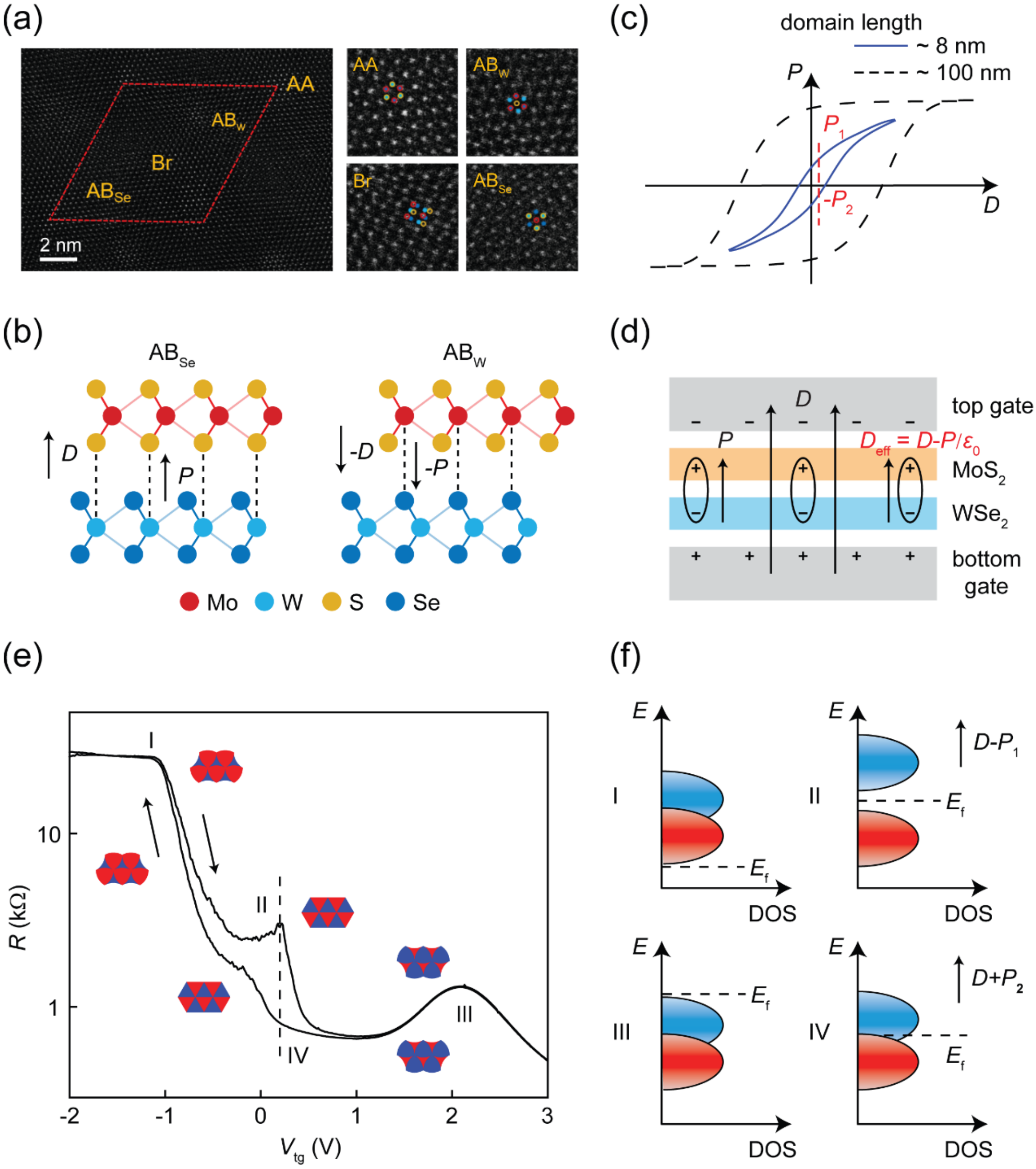

Fig. 4. (a) TEM image showing high-symmetry stacking regions: AA (atom-on-atom), $AB_{Se}$ (S-on-W), and $AB_W$ (Mo-on-Se), with Br bridging $AB_{Se}$/$AB_W$. A triangular moiré pattern (period ~8 nm) is visible. (b) Schematic illustration of the interfacial polarization under strong upward/downward external *D* field. (c) Schematic *P*-*D* hysteresis loop. Polarization switches between $+P_1$ and $-P_2$ at certain *D* during sweeps. (d) Schematic of the dipole and field configuration in $MoS_2$/$WSe_2$ heterobilayer. The interfacial polarization enabling effective field modulation via $D_{\mathrm{eff}} = D - P/\epsilon_0$. (e) Resistance hysteresis and phase transition at $V_{\mathrm{bg}} = 73$ V. The black arrows indicate the sweep direction. (f) Schematic band structure: upper/lower Hubbard bands (blue/red) and Fermi level (dashed lines). The Hubbard band gap is larger for forward sweep (II, smaller field) and smaller for backward sweep (IV, larger field).